\documentclass[
aps,%
12pt,%
final,%
notitlepage,%
oneside,%
onecolumn,%
nobibnotes,%
nofootinbib,% 
superscriptaddress,%
noshowpacs,%
centertags]%
{revtex4-2}

\usepackage{hyperref}

\begin{document}

\title{Spin Physics at NICA SPD}

\author{\firstname{I.}~\surname{Denisenko}\\
(on behalf of the SPD Collaboration)}
\email{iden@jinr.ru}
\affiliation{%
Joint Institute for Nuclear Research
}%

%\date{\today}
%\today печатает cегодняшнее число

\begin{abstract}
The Spin Physics Detector (SPD) is a universal detector at the NICA collider under commissioning at JINR, Dubna. The SPD is intended to study the spin structure of the proton and deuteron and other spin-related phenomena using a unique possibility to operate with polarized proton and deuteron beams at a collision energies up to 27~GeV and a luminosity up to 10$^{32}$cm$^{-2}$s$^{-1}$. As the primary goal, the experiment aims to provide access to the gluon TMD PDFs in the proton and deuteron, as well as the gluon transversity distribution and tensor PDFs in the deuteron, via the measurements of relevant single and double spin asymmetries using different complementary probes such as charmonia, open charm, and prompt photon production processes. Other polarized and unpolarized physics studies are possible, especially at the first stage of NICA operation with reduced luminosity and collision energy of the proton and ion beams. The physics program of the SPD and the design of the SPD setup will be presented.
\end{abstract}

\maketitle

\section{Introduction}
The primary goal of the SPD (Spin Physics Detector) experiment being constructed
at the Nuclotron-based Ion Collider Facility (NICA) is to provide a
comprehensive studies of the polarized and unpolarized gluon structure of the proton and
deuteron. For such studies, NICA is expected to provide
symmetric proton-proton and deuteron-deuteron collisions at $\sqrt{s} = 27$~GeV
and $\sqrt{s_{NN}} = 13.5$~GeV, respectively. The beam polarization during the run is expected to be
not less than 70\%. The designed maximum luminosity is $10^{32}$cm$^{-2}$s$^{-1}$
for $pp$ collisions and
$10^{31}$cm$^{-2}$s$^{-1}$ for $dd$ collisions
at the maximum energy. As a facility with polarized proton beams, NICA
covers a unique energy range
between COSY, SATURNE~II, and  U-70 at lower energies and RHIC (and proposed
experiments LHCSpin and AFTER at LHC) at higher collision energies.

For SPD, three complementary processes have been selected as the main probes of nucleon
gluon structure at NICA energies: charmonium production, open charm production,
and prompt photon production. Measurements of these processes cover high $x$
(the Bjorken variable) and a relatively low energy scale.

The project will be implemented in two phases. During the first phase,
only part of the detector subsystems will be installed, and collider
will operate at reduced energy and luminosity.

\section{Detector}
\begin{figure}
	\setcaptionmargin{5mm}
	\onelinecaptionstrue
	\includegraphics[width=.8\textwidth]{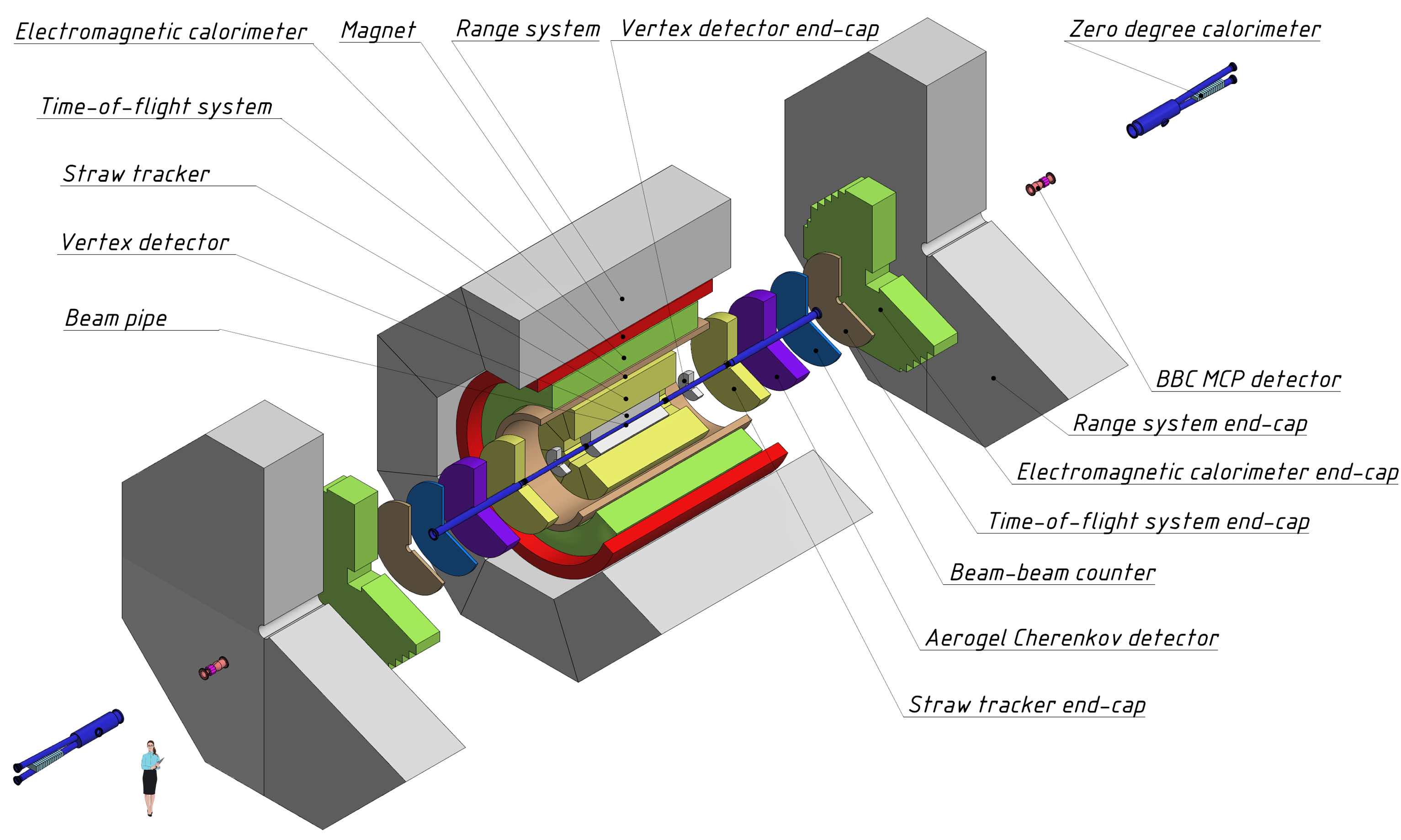}
	\captionstyle{normal}
	\caption{SPD detector schematic view for Phase~II.}
	\label{fig:SPD-set-up}
\end{figure}

The detector is optimized for the measurements  of charmonia, prompt photons
and open charm production processes. Its detailed description
can be found in the TDR~\cite{SPD:2024gkq}. A schematic view of the detector
is shown in Figure~\ref{fig:SPD-set-up}. The subsystem closest to the beam pipe
is silicon vertex detector, aimed at the
reconstruction of $D$-meson decays, which will be built
based on MAPS or DSSD technology. Next is the straw tracker
(ST) with spatial resolution of 150~$\mu m$. Particle
identification will be performed via ionization losses in
ST (with resolution of 8.5\%), time-of-flight system (TOF) with
a time resolution of 50~ps, and FARICH detectors in the endcaps
to provide $3\sigma$ $\pi/K$ separation up to $p=5.5$~GeV.
Photons will be detected by the electromagnetic calorimeter (ECal)
with a relative resolution of $5\%/\sqrt{E}\oplus1\%$.
The superconducting magnet will create magnetic
field up to 1.2~T in the central region of the detector. Muon identification
will be performed by the range system (RS). This system  also serves as
supporting frame for the whole detector and flux-return yoke for the magnet.
The luminosity monitoring and online polarimetry will be performed
by beam-beam counters and zero-degree calorimeters.

For the first phase, FARICH, TOF, and ECal will not be available. Instead of
silicon vertex detector a cheaper three-layer central tracker built using the
Micromegas technology will be installed.

To reduce possible bias, the detector will operate in triggerless mode. Nevertheless,
to make the system computationally manageable, the input data stream will be reduced by
a factor 20--50 by the software trigger (online filter).

% FIGURES: scheme of the experiment

\section{SPD Physics Program}
The physics program is naturally divided into two parts corresponding to the first
and second phases of the experiment. The first phase will take about two years
and cover the range $3.5\text{ GeV} < \sqrt{s} < 9.4\text{ GeV}$ for polarized $pp$ collisions.
Deuterons and light nuclei will be accelerated up to a kinetic energy of 4.5~GeV/A.
For the second phase, which will take about four years, the $pp$ collision energy is
$10\text{ GeV} < \sqrt{s} < 27\text{ GeV}$, and the detector will be assembled to its
full configuration.

\subsection{Phase I}
The Phase~I energy range is the transition region between the
hadronic and quark-gluon degrees of freedom. A variety of aspects of QCD can be
 studied in polarized and unpolarized nucleon collisions. These include studies of
 spin effects in elastic $pp$ and $dd$ scattering;
 asymmetries in hadron production, hyperon polarization, and vector
 meson alignment in collisions  of polarized beams;
 spin correlations in hyperon pair production;
 fluctons in $dd$ collisions;
 diquarks via high-$p_T$ meson and baryon production;
 quark-instanton scattering via measuring the dependence of hyperon polarization
  on collision energy;
 non-baryonic content of the deuteron in high-angle deuteron scattering;
 color transparency in quasi-free $pd$ interactions;
 hypernuclei productions (including the search for the doubly strange ones);
 physics of light and moderate ion collisions,
 among others.
Some of these topics are discussed in detail in Refs.~\cite{Abramov:2021vtu,SPDproto:2021hnm}.

\subsection{Phase II}
Understanding how hadrons are formed and how their properties like mass and spin emerge
the from the fundamental
degrees of freedom in QCD---quarks and gluons---is one of the key questions of particle
physics. One of the approaches to address this problem
is the study of parton distributions in both the transverse momentum ($\bf k_T$) space, via
transverse-momentum-dependent distributions (TMDs), and in the transverse position space,
via generalized parton distributions (GPDs).
TMDs can be studied via inclusive hadron production, whereas the extraction of GPDs requires
measurements of exclusive processes, which are largely not feasible at the SPD and,
consequently, are not discussed in this work.

Substantial knowledge has already been accumulated for unpolarized and polarized quark PDFs
in the proton (for a comprehensive review of quark TMDs, see Ref.~\cite{Boussarie:2023izj}).
For gluons, however, situation is quite different.
Although the unpolarized collinear gluon density, $f^g(x)$, is well established,
its uncertainties at high $x$ remains sizable. There has been substantial progress
on the collinear helicity PDF, $\Delta g(x)$, which describes PDF modification
for a longitudinally polarized proton
(see Ref.~\cite{Cruz-Martinez:2025ahf} and references therein).
%Bertone:2024taw, Borsa:2024mss}
The leading-twist gluon TMDs are shown in Table~\ref{tab:gluon_TMDs}.
%Notably the
%number of independent functions is in fact doubled by the presence of distinct
%gauge-link structures.
Upon $\bf{k}_T$-integration for a spin 1/2 particle
$f_1^g(x, \mathbf{k}_T^2)$ is reduced to $f^g(x)$, $g_1^g(x, \mathbf{k}_T^2)$ is reduced to
$\Delta g(x)$, the remaining functions vanish.
The distributions relevant to the SPD physics program are
the unpolarized gluon distribution,
$f_1^g(x, \mathbf{k}_T^2)$; the gluon Sivers
function (GSF), $f_{1T}^{\perp g}(x, \mathbf{k}_T^2)$, which encodes the spin--momentum correlation in
a transversely polarized proton; and the gluon Boer-Mulders function,
$h_1^{\perp g}(x, \mathbf{k}_T^2)$, which describes the linear polarization of gluons inside
an unpolarized hadron.
For the deuteron, it is of particular interest to probe its transversity
distribution, $h_1^g(x)$: contrary to the spin-1/2 hadron case, this function may be non-zero,
which would indicate the presence of non-nucleonic degrees of freedom.
There have been attempts to probe $f_1^g(x, \mathbf{k}_T^2)$ and $h_1^{\perp g}(x, \mathbf{k}_T^2)$,
including a recent analysis of LHC data on Higgs boson production~~\cite{Anedda:2026cox}.
For the $f_{1T}^{\perp g}(x, \mathbf{k}_T^2)$ function, there were phenomenological studies
based inclusive $\pi^0$ and $D$-meson azimuthal asymmetries from PHENIX
data~\cite{DAlesio:2015fwo,DAlesio:2018rnv}. Overall, gluon TMDs remain largely unconstrained.

 \begin{table}
	\setcaptionmargin{0mm} \onelinecaptionsfalse
\captionstyle{flushleft}
	\caption{Nucleon gluon TMD PDFs at twist-2 (adapted from Ref.~\cite{Arbuzov:2020cqg}).
	}
\label{tab:gluon_TMDs}
	\hspace{1cm} gluon pol. \\ \vspace{0.1cm}
	\rotatebox{90}{\hspace{-1cm} nucleon pol.} \hspace{0.1cm}
	%\begin{tabular}[c]{|m{0.7cm}|c|c|c|}  % for older Latex version
	\begin{tabular}[c]{|w{c}{0.7cm}|c|c|c|}
		\hline
		& $U$ & circular & linear \\
		\hline
		$U$ & $f_{1}^{g}$ & & $h_{1}^{\perp g}$ \\
		\hline	
		$L$ & & $g_{1}^{g}$ & $h_{1L}^{\perp g}$ \\
		\hline	
		$T$ & $f_{1T}^{\perp g}$ & $g_{1T}^{g}$ & $h_{1}^{g}$, $h_{1T}^{\perp g}$ \\
		\hline
	\end{tabular}
\end{table}

At SPD, we will be able to access both the collinear PDFs, $f^g(x)$
and $\Delta g(x)$, and the mentioned TMDs via measurements of three complementary
physics processes: charmonium production, open charm production,
and prompt photon production (a comprehensive discussion of proposed studies
is presented in Ref.~\cite{Arbuzov:2020cqg}).
Such studies constitute the core of the SPD Phase~II physics program.
Among the most interesting observables are those arising in collisions
of polarized beams. Longitudinal double asymmetries, $A_{LL}$,
are sensitive to parton helicity distributions.
Azimuthal modulations in particle production from collisions of transversely polarized
and unpolarized beams (single-spin asymmetries), $A_N$, probe the Sivers
function.

\subsubsection{Measurements with charmonia}
The description of charmonium production at SPD is model-dependent.
Usually, two models (or their modifications) for $c\bar c$ quark
pair hadronization to charmonium are used: Non-relativistic QCD (NRQCD)
and the Color Evaporation Model (CEM).
For all models at SPD Phase~II energies charmonium production
is dominated by the gluon-gluon fusion. The difference in predictions
between these models complicates the interpretation of the experimental
results and requires comprehensive model validation. Moreover,
inclusive $J/\psi$ production is a TMD
factorization breaking process, and the size of this violation should
be studied experimentally.

\begin{figure}
	\setcaptionmargin{5mm}
	\onelinecaptionstrue
	\begin{tabular}{c @{\hspace{2em}} c}
		\includegraphics[width=0.39\textwidth, height=0.27\textwidth]{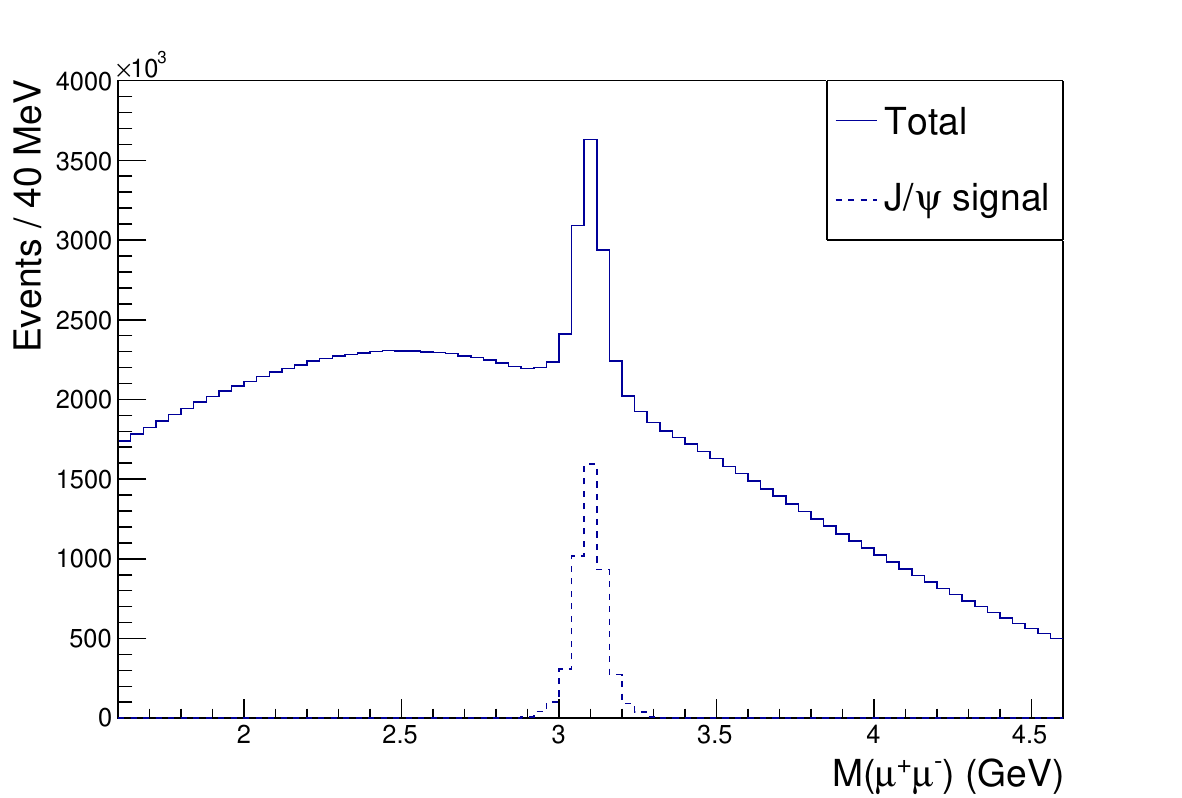} &
		\includegraphics[width=0.5\textwidth, height=0.27\textwidth]{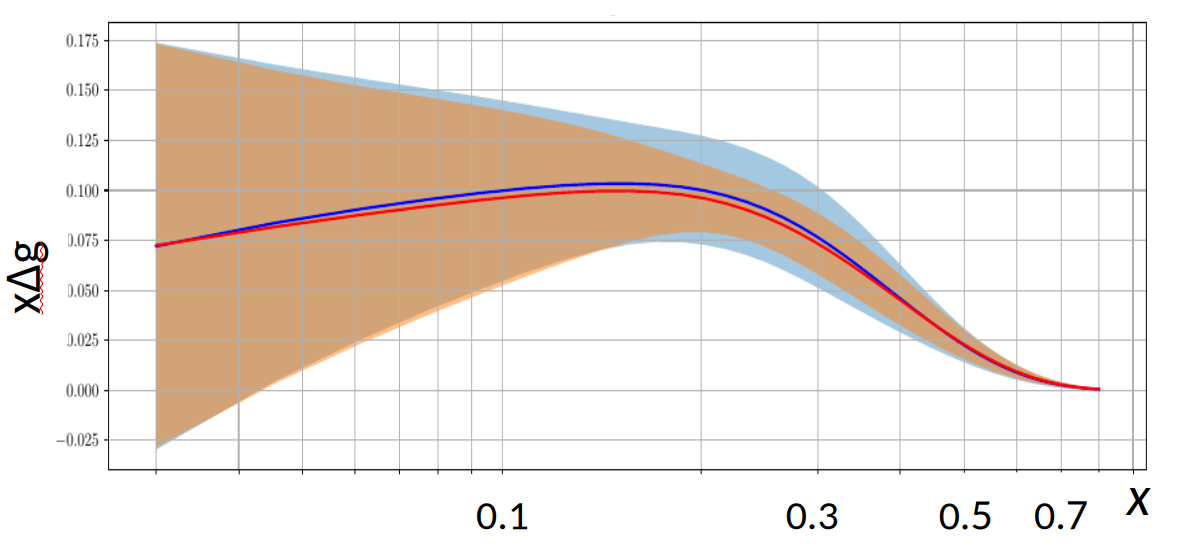} \\
		\small (a) & \small (b)
	\end{tabular}
	\captionstyle{normal}
	\caption{(a) Invariant mass distribution of $\mu^+\mu^-$ candidate pairs.
		    (b) Estimated impact of SPD $J/\psi$ $A_{LL}$ measurements on the $x\Delta g$ uncertainty
			using the NNPDFpol1.1 set of replicas.
			The PDF uncertainty bands without and with SPD pseudodata are shown in blue and brown, respectively.
			}
	\label{fig:jpsi}
\end{figure}

The inclusive $J/\psi$ production has a large
cross-section (200--250~nb at the SPD maximum energy) and a clear
experimental signature in the $J/\psi\to\mu^+\mu^-$ decay. Muons can
be identified via patterns and track segment length in the RS. For a
very rough selection, the spectrum of muon candidate pairs is
shown in Figure~\ref{fig:jpsi}(a) (for details, see Ref.~\cite{SPDproto:2021hnm}).
The relatively small beam collision
energies at SPD result in soft charged pion momentum spectra and
thus larger background due to their decays and misidentification.
Overall, we expect about 5 million reconstructed $J/\psi\to\mu^+\mu^-$
events per year (here and in the following, one year corresponds to
$10^7$~s of data-taking time) at the maximum collision energy.
Despite the significant background, we expect precise measurements
of differential cross-sections, polarization, and asymmetries.

The PHENIX collaboration measured $J/\psi$ $A_{LL}$ at 510~GeV
in a narrow kinematic region~\cite{PHENIX:2016gzj}.
At the SPD, effective $x$ of interacting gluons are in the region
where $\Delta g$ is the largest, hence asymmetries of the level
1--10\% are expected; the acceptance covers a wider rapidity region |y| < 2,
and the statistical uncertainties are estimated to be smaller.
The impact of SPD $J/\psi$ measurement was determined
by estimating the $A_{LL}$ statistical uncertainties per year of data taking
(systematic uncertainties are ignored),
generating pseudodata, and using the Bayesian reweighting method
for Monte Carlo replicas to evaluate new PDF uncertainties.
The last two steps used the NNPDFpol1.1 set~\cite{Nocera:2014gqa} of helicity PDF replicas.
The PDF uncertainty bands before and after taking into account
the SPD pseudodata are shown in Figure~\ref{fig:jpsi}(b). For $x$ in the
range 0.2--0.3, the uncertainty reduces by a factor of 2.

\begin{figure}
	\setcaptionmargin{5mm}
	\onelinecaptionstrue
	\begin{tabular}{c @{\hspace{2em}} c}
		\includegraphics[width=0.4\textwidth, height=0.25\textwidth]{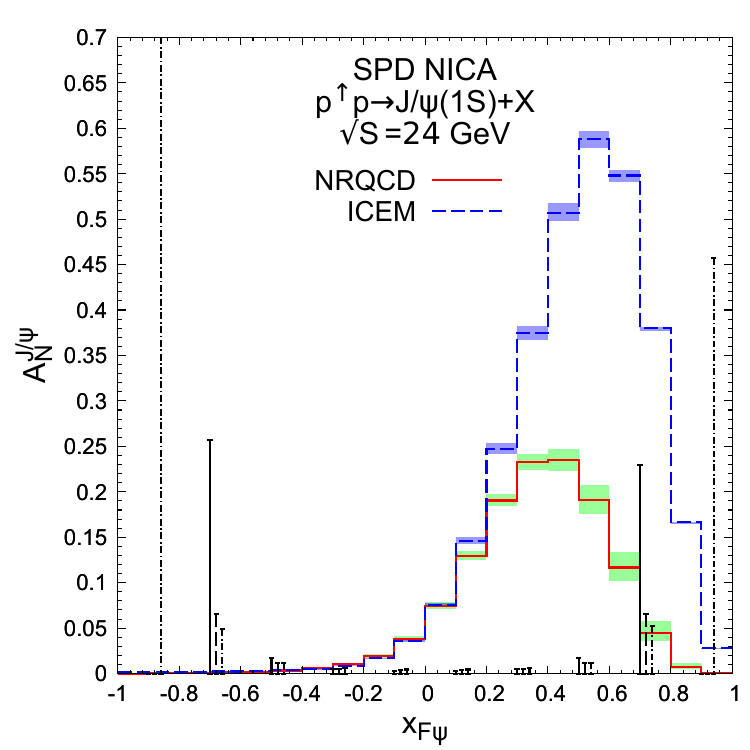} &
		\includegraphics[width=0.4\textwidth, height=0.25\textwidth]{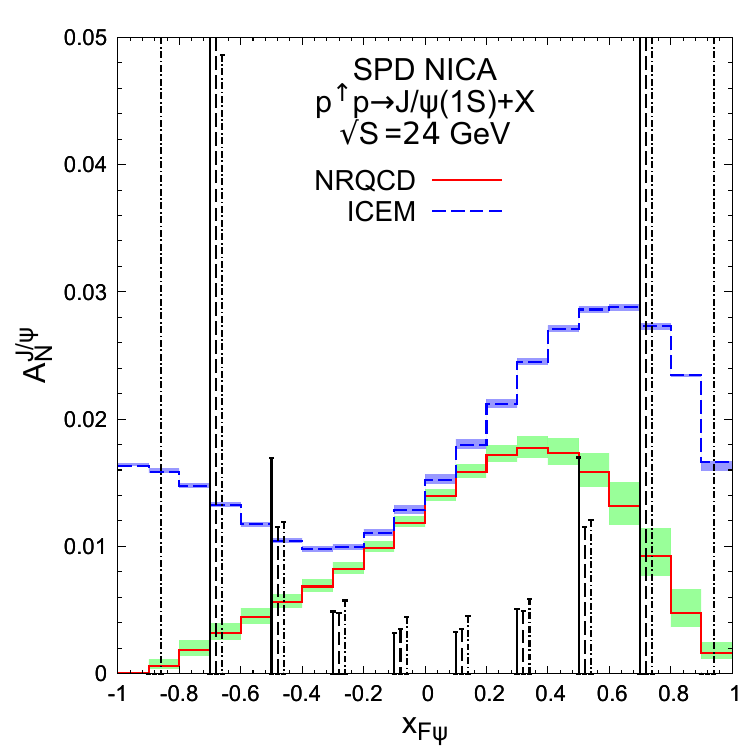} \\
		\small (a) & \small (b)
	\end{tabular}
	\captionstyle{normal}
	\caption{Comparison between the SPD statistical uncertainties (one year
		of data taking) and the GPM prediction for two different extractions of GSF~\cite{Karpishkov:2020brv}: (a) SIDIS1 and (b) D'Alesio.}
			\label{fig:jpsi-AN}
\end{figure}

The transverse single spin asymmetries for $J/\psi$ production were measured
by PHENIX~\cite{PHENIX:2010hqq,PHENIX:2018qvl} at a $pp$ collision energy of
200~GeV. The COMPASS collaboration also reported preliminary measurements of
such asymmetry in pion-induced $J/\psi$ production on a polarized target.
Both results were obtained in the limited $x_F$ range ($x_F = 2p_L^{J/\psi}/\sqrt{s}$) and are consistent
with zero. For the SPD, the statistical precisions of such measurements were estimated and is compared to the GPM
prediction~\cite{Karpishkov:2020brv} for two phenomenological extractions
of the gluon Sivers function in Fig.~\ref{fig:jpsi-AN}. The CGI-GPM
in the same publication predicts a lower asymmetry.

\begin{figure}
	\setcaptionmargin{5mm}
	\onelinecaptionstrue
		\begin{tabular}{c @{\hspace{2em}} c}
		\includegraphics[width=0.45\textwidth, height=0.24\textwidth]{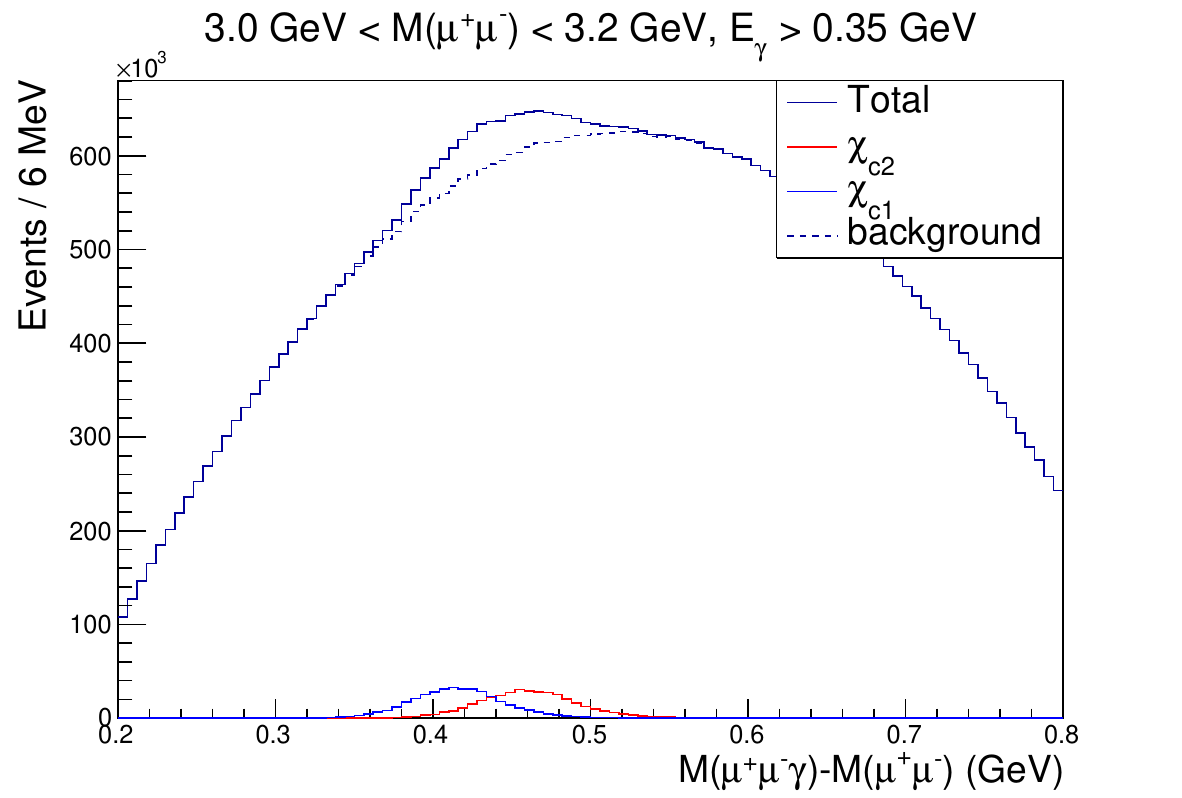} &
		\includegraphics[width=0.45\textwidth, height=0.24\textwidth]{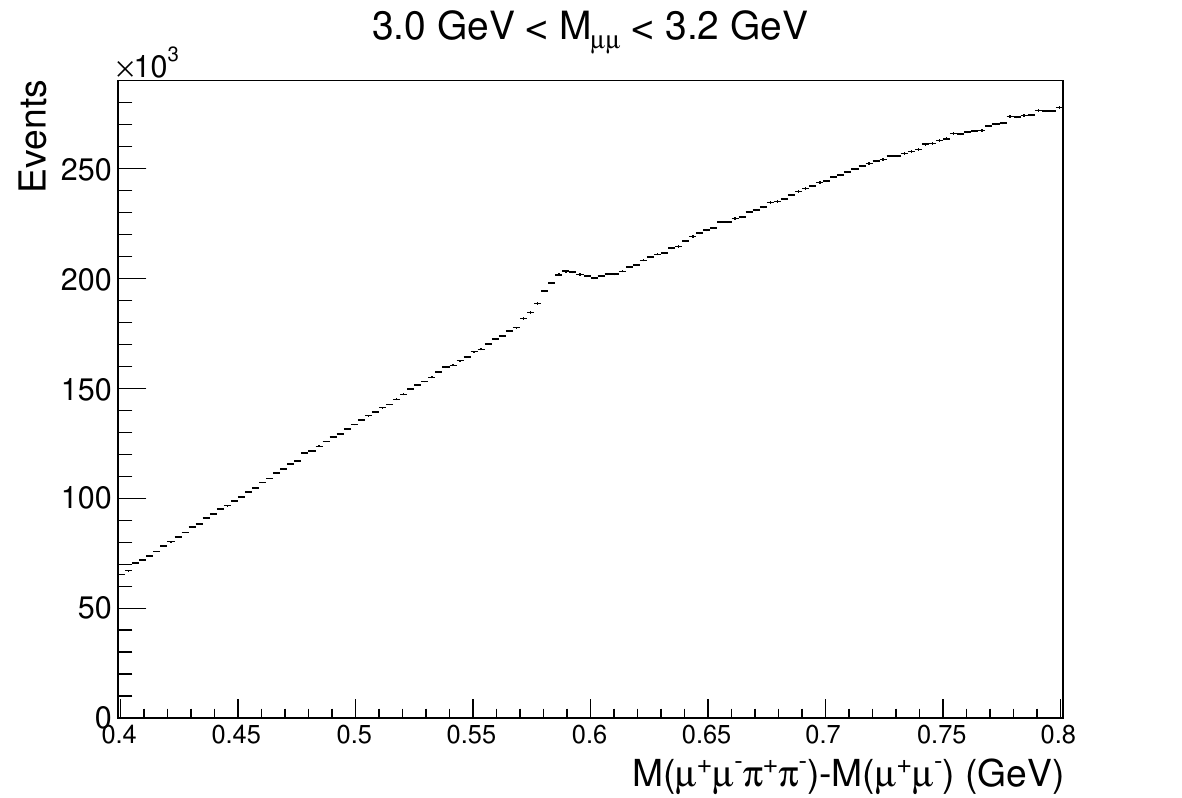} \\
		\small (a) & \small (b)
	\end{tabular}
	\captionstyle{normal}
	\caption{(a) $\chi_{c1}$ and $\chi_{c2}$ signals and estimated background. (b)
		The $\psi(3686)$ signal reconstructed from its decay to $J/\psi\pi^+\pi^-$. Both
		plots are estimates for one year of detector operation.}
	\label{fig:chi-psi2S}
\end{figure}

Unlike the previous beam-dump experiments at the same energies, SPD
can measure the production of heavier charmonia states decaying to $J/\psi$
($\chi_{c1}$, $\chi_{c2}$, and $\psi(3686)$) and determine the
feed-down contributions from their decays. The $\chi_{c1}$ and $\chi_{c2}$
states can be reconstructed from their radiative decays to $J/\psi$.
The distribution of the $M_{J/\psi\gamma} - M_{J/\psi}$ variable is shown
in Figure.~\ref{fig:chi-psi2S}(a). For both states together, we expect
about 0.5 million selected events per year against a very strong background.
The ECal energy resolution is not
sufficient to separate $\chi_{c1}$ and $\chi_{c2}$, but measurement of
the relative contribution of each state should be possible. In
Figure.~\ref{fig:chi-psi2S}(b) the $\psi(3686)\to J/\psi\pi^+\pi^-$ signal
is shown, approximately 0.1 million selected events are expected per year.

Studies of associated charmonium production ($J/\psi J/\psi$, $\gamma J/\psi$, and others)
and exclusive $J/\psi$ production~\cite{Xie:2025sfx} are also considered for the SPD.

\subsubsection{Measurements with prompt photons}
\begin{figure}
	\setcaptionmargin{5mm}
	\onelinecaptionstrue
	\begin{tabular}{c @{\hspace{2em}} c}
		\includegraphics[width=0.45\textwidth, height=0.26\textwidth]{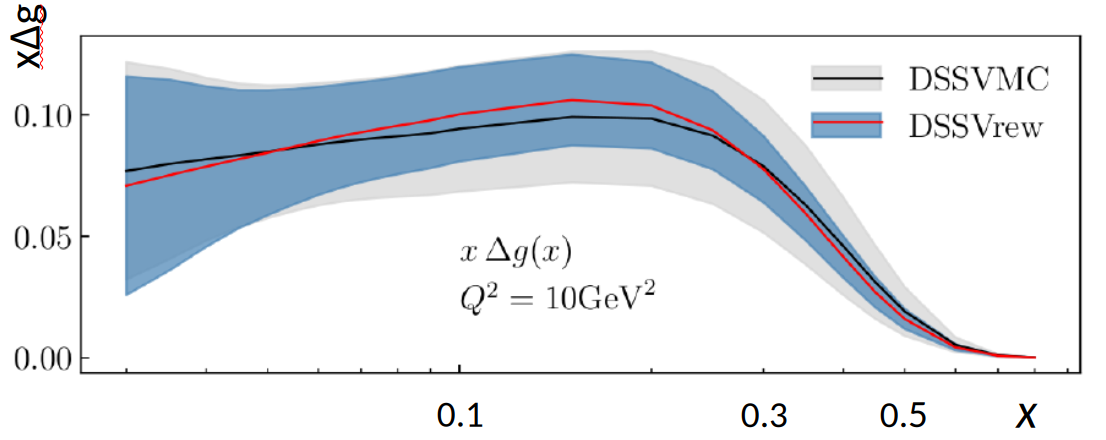} &
		\includegraphics[width=0.45\textwidth, height=0.25\textwidth]{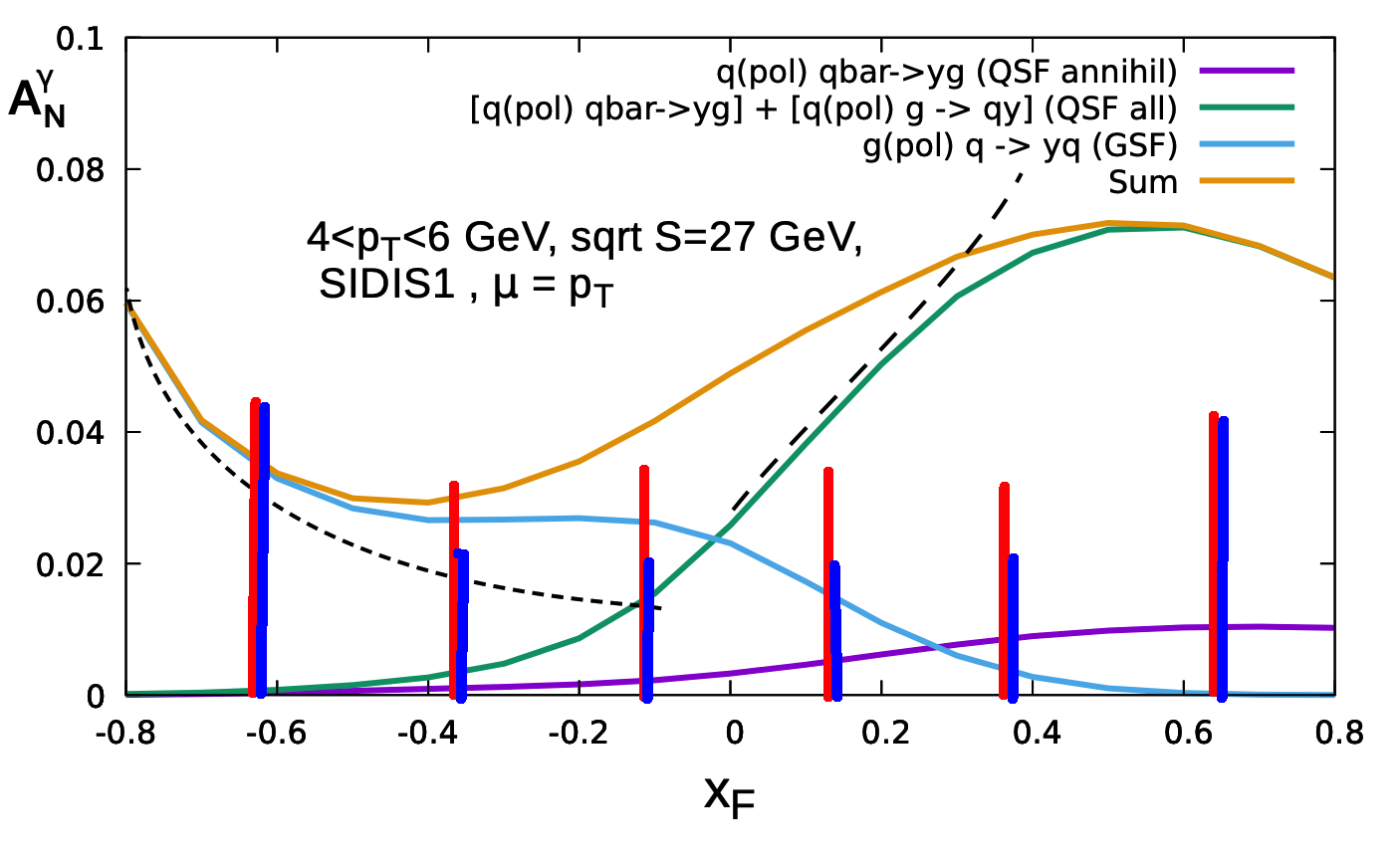} \\
		\small (a) & \small (b)
	\end{tabular}
	\captionstyle{normal}
	\caption{
		(a) Estimated impact of SPD prompt photon $A_{LL}$ measurements on the $x\Delta g$
		uncertainty, calculated using the DSSV14 set of replicas. The PDF uncertainty
		band without SPD pseudodata is shown in gray, while the band with SPD
		measurements is shown in blue (Courtesy of W.~Vogelsang, R.~Sassot, and I.~Borsa).		
		(b) Predicted $A_N$ for prompt photons at SPD compared to the estimated
		measurement uncertainty~\cite{SPDproto:2021hnm}.}
	\label{fig:prompt-photons}
\end{figure}

Theoretically, prompt photon production is the cleanest probe of gluon structure.
At the same time, measurements of this process are extremely challenging experimentally:
the process has a small cross-section and a huge background dominated by hadron decays.
Estimations show that the compromise between statistical and the expected systematic uncertainties
is to measure $p_T^\gamma > 4$~GeV/$c$.  The expected impact of SPD $A_{LL}$ measurements is shown
in Figure.~\ref{fig:prompt-photons}(a): the DSSV14 uncertainties are reduced by a factor of~2
in the high $x_F$ region. The transverse single spin asymmetry was measured before by
E704~\cite{E704:1995her} and PHENIX~\cite{PHENIX:2021irw}; both results are consistent
with zero. The SPD sensitivity to prompt
photon $A_N$ and contribution of different partonic subprocess is shown in Figure.~\ref{fig:prompt-photons}(b), where the contribution from the GSF dominates at negative $x_F$.

\subsubsection{Measurements with open charm}
\begin{figure}
	\setcaptionmargin{5mm}
	\onelinecaptionstrue
	\begin{tabular}{c @{\hspace{2em}} c}
		\includegraphics[width=0.4\textwidth, height=0.25\textwidth]{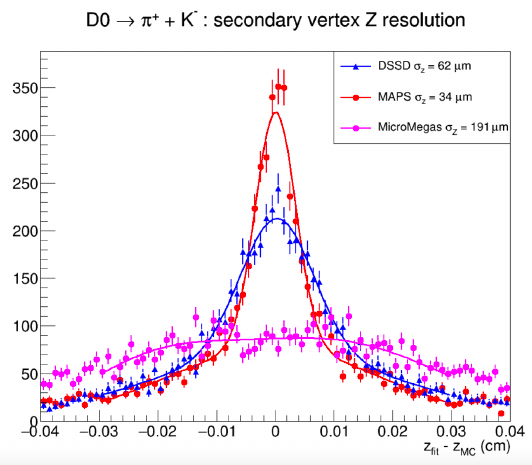} &
		\includegraphics[width=0.45\textwidth, height=0.25\textwidth]{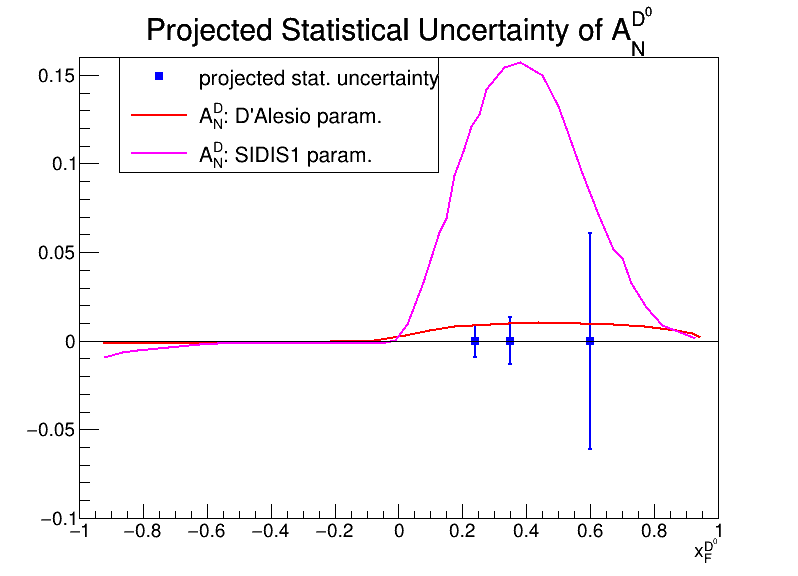} \\
		\small (a) & \small (b)
	\end{tabular}
	\captionstyle{normal}
	\caption{(a) SPD $\sigma_z$ resolution for $D^0$ vertex reconstruction.
		     (b) SPD $A_N$ statistical uncertainties compared to two model
		     predictions~\cite{Arbuzov:2020cqg}.
		    }
	\label{fig:D-mesons}
\end{figure}

$D$-mesons at the SPD can be reconstructed via decays to $K\pi$ ($D^0/\bar D^0$) and $K\pi\pi$
($D^\pm$). The major experimental difficulty is the small boost at SPD energies resulting in a strong
combinatorial background. At the same time, the large process cross-section allows the application of aggressive cut strategy. The SPD vertex detector resolution is shown in Figure~\ref{fig:D-mesons}(a).

Open charm production is suggested for precise measurement of $g(x)$ at large $x$,
as well as gluon helicity and Sivers functions (see Figure~\ref{fig:D-mesons}(b));
the $D$-meson pair production
probes the Boer-Mulders function. In all cases, the interpretation of SPD results
relies on precise knowledge of polarized and unpolarized $D$-meson fragmentation functions.

\subsection{Deuteron gluon structure}
Polarized deuteron beams open up the possibilities of studying non-nucleonic degrees of freedom
in the deuteron via measurements of gluon transversity and the tensor-polarized gluon distribution,
as well as modifications of the unpolarized PDF using the discussed probes
(for discussion, see Ref.~\cite{Arbuzov:2020cqg}). 

\section{Summary}
The SPD experiment is a comprehensive facility for studying the polarized and unpolarized
gluon structure of the proton and the deuteron in a unique energy region. The detector is
optimized for the charmonium, prompt photon, and open charm production measurements.
The expected results will improve our understanding of collinear PDFs and TMDs.
Alongside flagship measurements, a variety of aspects of QCD in the polarized
collisions can be investigated, especially during the first phase of the experiment.

The physics program of the SPD experiment with respect to nucleon gluon content is
complementary to those of experiments at RHIC and EIC, the proposed fixed target program
at LHC (AFTER, LHC-Spin), and EicC.

\bibliography{bibl}

\end{document}